# Multi-armed Bandit Learning for TDMA Transmission Slot Scheduling and Defragmentation for Improved Bandwidth Usage

Hrishikesh Dutta, Amit Kumar Bhuyan, and Subir Biswas
Michigan State University, East Lansing, USA
duttahr1@msu.edu, bhuyanam@msu.edu, sbiswas@egr.msu.edu

*Abstract* — **This paper proposes a Time Division Multiple Access (TDMA) MAC slot allocation protocol with efficient bandwidth usage in wireless sensor networks and Internet of Things (IoTs). The developed protocol has two primary components: a Multi-Armed Bandits (MAB)-based slot allocation mechanism for collision free transmission, and a Decentralized Defragmented Slot Backshift (DDSB) operation for improving bandwidth usage efficiency. The proposed framework is decentralized in that each node finds its transmission schedule independently without the control of any centralized arbitrator. The developed mechanism is suitable for networks with or without time synchronization, thus, making it suitable for low-complexity wireless transceivers for wireless sensor and IoT nodes. This framework is able to manage the trade-off between learning convergence time and bandwidth. In addition, it allows the nodes to adapt to topological changes while maintaining efficient bandwidth usage. The developed logic is tested for both fully-connected and arbitrary mesh networks with extensive simulation experiments. It is shown how the nodes can learn to select collision-free transmission slots using MAB. Moreover, the nodes learn to self-adjust their transmission schedules using a novel DDSB framework in order to reduce bandwidth usage.**

## I. INTRODUCTION

The primary objective of this work is to develop an online learning framework for TDMA slot allocation in wireless sensor and IoT networks. Multi-Armed Bandits (MAB) learning and a novel slot defragmentation operation are used in order to achieve this objective. The main limitation of the traditional TDMA MAC protocols is that these logics are pre-programmed based on heuristics and past experience of network designs. As a result, such protocols cannot adapt well to network and traffic dynamics and various kinds of heterogeneities. This leads to wastage of precious networking resources, including bandwidth and energy. Such phenomena are particularly harmful for IoT and sensor networks in which energy and other resource wastage can be operationally detrimental. In addition, the traditional TDMA slot scheduling usually relies on network time synchronization. Accurate time synchronization among wireless networks nodes can be expensive to realize, especially in low-cost nodes with limited processing and communication resources. Moreover, the MAC layer performance in such networks can be very sensitive to even slight perturbations in the quality of time synchronization [1]. In order to address these shortcomings, this paper leverages the on-the-fly learning abilities of MAB for developing a decentralized MAC protocol for TDMA slot allocation. And that is done without relying on network time synchronization.

The framework proposed in this paper has two distinct components: an MAB-based TDMA MAC slot allocator, and a Decentralized Defragmented Slot Backshift (DDSB) operator (Fig. 1). The goal of the first component is to make the nodes learn transmission schedules in a decentralized manner. Allocating slots in the absence of time synchronization can result in high bandwidth redundancy, especially if the underlying learning mechanism is made to converge fast [2]. This problem can be ameliorated using the novel DDSB operation (Stage 2 in Fig. 1). In addition to improving bandwidth utilization efficiency, this mechanism assists the underlying MAB learning to converge faster by allowing a larger TDMA frame size than the minimum required frame size. Apart from ensuring faster convergence and improving bandwidth usage, the DDSB operation helps the nodes adapt to topological changes. This is especially useful in scenarios of node failure, where the slots of the failed nodes remain unutilized and hence can result in poor bandwidth utilization. This is because of pre-allocated frame size in the traditional TDMA approaches. This shortcoming is overcome by the proposed DDSB mechanism.

A notable feature of the developed approach is that the framework is decentralized in that each node learns its transmission schedules independently without explicitly sharing the learning parameters with each other. This is specifically useful in partially connected networks, where the nodes have limited network information visibility. This also makes the framework scalable with network size since the learning is done independently in each node, and its performance depends on network degree rather than the network size. Moreover, decentralized learning is computationally more efficient as compared to centralized learning in which a centralized agent, with access to complete network level information, learns optimal node behavior and downloads it to individual nodes [3]. Such centralized learning also typically requires additional network resources in terms a separate channel to download the learned policies to the nodes.

This paper has the following scopes and contributions. First, an MAB-based learning framework is proposed for TDMA slot allocation that works both with and without the presence of network time synchronization. Second, a novel decentralized defragmented slot backshift (DDSB) mechanism is developed to reduce bandwidth redundancy and improve bandwidth usage efficiency in wireless networks. The proposed approach is decentralized in that it does not require global network information and is designed to adapt to dynamic network topologies. Third, the trade-off between convergence time and bandwidth usage efficiency is studied and it is shown how DDSB operation can help manage this trade-off. Finally, with extensive simulation experiments, the proposed mechanism is functionally validated, and performance evaluated for generalized networks with arbitrary mesh topologies.

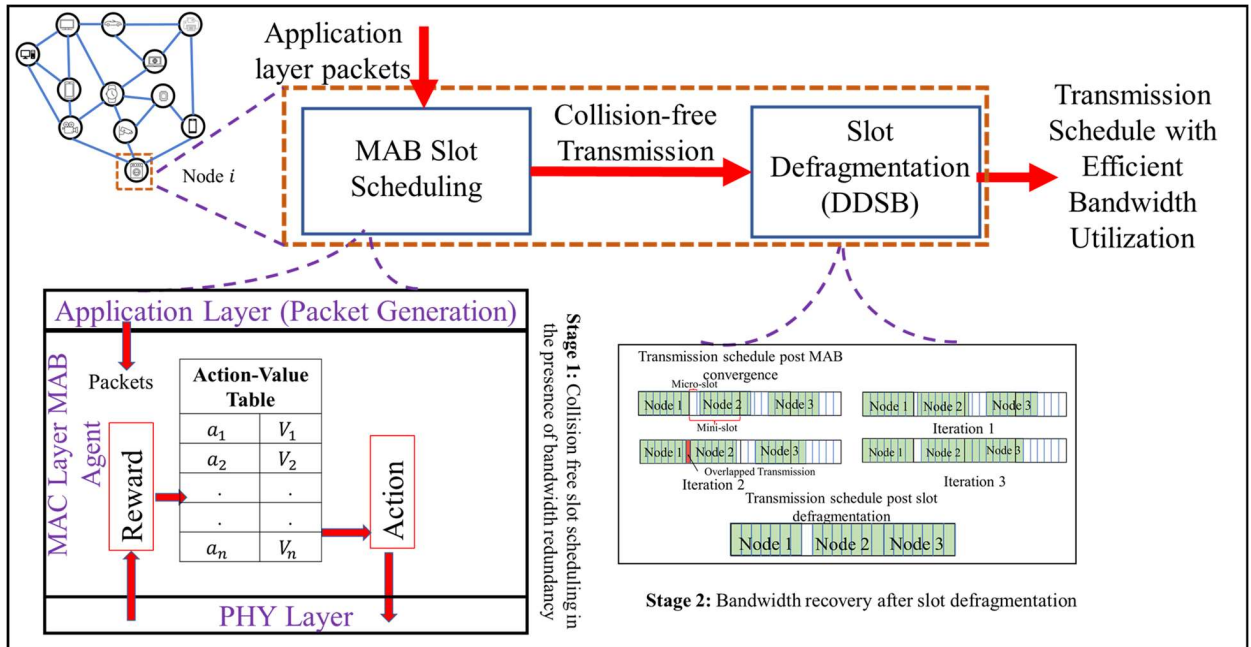

Fig. 1: High-level working model of the proposed scheduling framework with efficient bandwidth usage

## II. RELATED WORK

TDMA slot allocation in wireless networks using Reinforcement Learning and Multi-Armed Bandits have been explored in the literature. The authors in [10] propose an RL-based approach for allocating MAC slots and this contention-based technique achieves throughput which is higher than that of the standard protocols. However, this performance improvement does not hold for different network topologies. The mechanism proposed in [11] aims at reducing energy consumption using a centralized learning approach, which is not practical in many large networks that do not support centralized access arbitration. The papers in [12] use distributed Reinforcement Learning for enabling the nodes to learn an efficient sleep-awake schedule for minimizing energy consumption. However, these mechanisms rely on network time synchronization capability. Also, these protocols cannot adapt to changes in the network topology that would lead to spectral usage inefficiency in case of node failure.

There are papers that use distributed approaches for MAC slot allocations. The work in [12], has developed a contention-based mechanism for slot allocation fairly in wireless mesh networks. The authors in [13] propose a distributed framework for TDMA slot allocation in networks that use two-hops neighbor information. Unlike the framework used in this work, these mechanisms require a time synchronized network and lack adaptability to network topology changes.

There exist approaches for tackling the slot-assignment problem in time-asynchronous networks. The mechanism in [6] depends on a centralized gateway for transmission scheduling. The authors in [14] propose a distributed approach to schedule in the absence of global time synchronization. This framework requires the network nodes to explicitly share information with two-hop neighbors and the performance is heavily affected by the local clock drifts in the network. Also, it is shown in [7] that assigning slots in networks without time synchronization leads to bandwidth wastage.

In this work a decentralized framework is proposed to address the limitations of the existing developments in this field, where the nodes learn transmission policies independently and adaptively in networks with and without time synchronization.

## III. IoT NETWORK AND TRAFFIC MODEL

The proposed mechanism is developed for generalized point to point networks with arbitrary mesh topologies (i.e., fully connected and partially connected) and network traffic patterns. From a learning standpoint, the main difference between the two connectivity modes is the amount of slot allocation information availability at each node. For the fully connected case, each network node possesses the current MAC slot information for all other nodes in the network and for the latter, only localized information is available to the nodes.

The MAC layer traffic load model is created such that a packet generated in a node (using constant bit rate traffic model) is sent to one of its 1-hop neighbors chosen using uniform random distribution. In other words, if a node $i$ has $X$ one-hop neighbors and its MAC layer load is $\lambda_i$ packet per frame (ppf), node $i$ statistically sends $\frac{\lambda_i}{X}$ ppf amount of traffic to each of its neighbors.

The target mechanism would work both in the presence and absence of time synchronization in the network. This is a crucial feature since MAC slot allocation in the absence of time synchronization is a challenging problem, and it is a notable feature of the proposed learning mechanism in this paper. The network model includes the availability of piggybacking for sending control information using a small part of the data packets that allows the framework to be not dependent on the abilities of direct collision detection, which is especially meaningful for the low-complexity wireless transceivers in IoT/Sensor nodes [15].

## IV. TDMA SLOT ALLOCATION USING MAB

Multi-Armed Bandits (MAB) is a special class of Reinforcement Learning in a non-associative setting [16]. A much-explored variant of MAB is the '$k$-armed bandit' problem, where the learning agent (bandit) has $k$ possible arms or possible actions to choose from. Each of the $k$ actions has an associated stochastic reward the distribution of which is not known to the learning agent. The agent's goal is to maximize the total accumulated reward over infinite time horizon by learning to estimate reward distribution of the possible actions.

### A. MAB for TDMA Slot Allocation in Time-Synchronized Network

MAC slot allocation problem here refers to each node being able to choose a slot at which the node can transmit in all subsequent frames without colliding with the transmissions from the other network nodes. Such collision-free slots should be selected without any centralized allocation. The selection policy is modeled as an MAB problem, where each node acts as an '$F$-armed bandit' ($F$ is the frame size). The action of the bandit is to select a slot, representing an arm, from an action pool of $F$ slots, which is preset based on network size/degree.

The environment here is the wireless network with which the nodes/agents interact through their actions of choosing transmission slots (i.e., the bandit arms). The reward associated with an action is formulated such that a node or an agent receives a penalty if it selects a slot that overlaps with transmissions from other nodes, leading to collisions. Conversely, an action is rewarded for a collision-free transmission. The reward function for node $i$ in decision epoch $t$ is formulated as:

$$R_i(t) = \begin{cases} +1, \text{ success} \\ -1, \text{ collision} \end{cases} \quad (1)$$

Using the actions and the reward function mentioned above, each learning agent (i.e., a node) learns a transmission policy to avoid collision in a distributed manner

### B. MAB for TDMA Slot Allocation in the absence of network Time-Synchronization

Accurate time synchronization among wireless networks nodes can be expensive to realize especially in low-cost IoT nodes with limited processing and communication resources. Also, MAC layer performance in such networks can be very sensitive to even slight perturbations in the quality of time synchronization. This section explains the TDMA slot allocation using MAB in the absence of time synchronization.

Like regular TDMA, the framework would work with fixed size frame abstraction. The main catch here is that the frames are not synchronized across the network and the concept of frame is totally local to a node. A node decides the time of start of its own frame, and the frame end time is decided based on the fixed frame duration. The node does not know about the start times of the other network nodes' frames. Within a frame, a node can schedule a packet transmission only in certain discrete time instances away from its frame start time. The intervals between those time instances are referred to as mini-slots, the duration of which is an integer submultiple of the packet duration, and is equal at all nodes. The details on TDMA operation in time-asynchronous networks can be found in [9].

Transmission scheduling problem in this context boils down for each node to be able to choose a mini-slot at which the node can transmit in all subsequent frames without colliding with the transmissions from the other network nodes. Such collision-free mini-slots is selected locally at each node in a fully distributed manner, and that is without any centralized allocation entities and network time synchronization. The selection policy is modeled as a Multi-Armed Bandit problem as discussed in *Subsection IV B*. The only difference here is that instead of slot selection, the node has to pick the collision-free transmission mini-slot [9]. The MAB slot allocation model has been shown in Fig. 1 (Stage 1).

### C. Limitations of MAB-based TDMA slot Allocation

As explained in the prior section, each node learns independently over time to find a collision-free transmission slot or mini-slot in a frame using Multi-armed Bandit. However, there are major scopes of improvement of the framework post MAB convergence as explained below.

<u>Bandwidth Usage Efficiency-Convergence Time Trade-off</u>: Any learning for mini-slot selection would require nodes to perform certain amount of iterative search for a collision free transmission mini-slot within its own frame. Since the targeted learning is distributed in that each node performs its own independent search, short term collisions and scheduling deadlocks can occur. This can be mitigated by making the frames longer than the absolutely minimum required length, leading to certain amount of bandwidth redundancy. This redundancy can be expressed by a factor $K = \frac{\text{Frame Size}}{\text{Minimum frame size}}$.

The minimum frame size in the denominator represents the absolute minimum frame size that is possible in the presence of time synchronization. To be noted that, and as will be shown in Section VI, the frame scaling factor $K$ plays a significant role in the convergence speed of the learning. For a given network, convergence is faster with increase in the value of $K$. This is because, with larger value of $K$, the number of feasible solutions of the MAB problem increases and hence the probability of finding a collision-free transmission strategy increases. However, increase in the frame scaling factor $K$ increases the frame size causing bandwidth usage inefficiency. In short, there exists a trade-off between the convergence time and bandwidth usage efficiency post MAB convergence.

<u>Bandwidth Redundancy in the absence of network time synchronization</u>: As shown in [10], in order to assign collision-free transmission mini-slots in networks without time synchronization capability, the minimum frame length should be at least one mini-slot more than the absolute minimum frame size. In other words, the frame-scaling factor ($K$) should be greater than 1. This leads to certain amount of bandwidth redundancy because of the extra mini-slot in the frame without any packet transmission. Higher the value of $K$, higher is the bandwidth redundancy in time-asynchronous network.

<u>Inadaptability to network topology changes</u>: In classical TDMA systems and also in this MAB-based (mini)slot allocation mechanism, the frame size is preset based on the network size/ degree and the frame scaling factor $K$. However, in dynamic network topologies, where nodes can enter and leave the network, having a preset frame size is not suitable. On one hand, a smaller frame size (new nodes entering the network) will lead to collisions and on the other, if the frame size is larger than required (nodes leaving the network), it would lead to bandwidth usage inefficiency as discussed above. While the former can be easily solved by making the nodes learn using the MAB framework proposed in Section IV B, as soon as they detect any collisions after steady state, the later problem requires special approach for solving, since there is no way for the nodes to know that there are nodes leaving the network.

In order to solve the problems explained above, we propose the Decentralized Defragmented Slot Backshift (DDSB) operation (Stage 2 in Fig. 1) for reducing bandwidth redundancy and increasing bandwidth usage efficiency.

V. DECENTRALIZED DEFRAGMENTED SLOT BACKSHIFT (DDSB)

In this section, a post-convergence distributed defragmented slot backshift (DDSB) mechanism is introduced to reduce capacity overhead, bandwidth usage inefficiency and bandwidth redundancy-convergence time trade-off. This step is performed by all the nodes independently after they have found a suitable (mini)slot for transmission post MAB convergence. This is also executed periodically in a dynamic topology to reduce any bandwidth usage inefficiency because of nodes leaving the network.

This concept is implemented by discretizing each slot or mini-slot within a frame into '$s$' number of micro-slots. After MAB convergence, each node shifts its packet transmission by one micro-slot back in time till it experiences a collision. Once a node experiences a collision it undoes its previous action to find its new transmission micro-slot. In this way, the nodes make an estimate of the unused space in the frame and try to reduce it in a distributed manner.

This DDSB operation is explained using Fig. 1 (Stage 2) for a 3-nodes fully connected network without time synchronization. The figure shows how the frame structure (with reference to node 1) evolves over 5 iterations of defragmentation mechanism for $K = 1.33$ and $s = 7$. In this figure, node 1 does not shift its transmission since it is transmitting at the beginning of the frame. Nodes 2 and 3 backshift their transmissions by one micro-slot per iteration. In iteration 2, nodes 1 and 2 experience a collision and hence node 2 undoes its previous action by shifting by two-micro-slot forward in iteration 3. But node 1 does nothing in iteration 3, since it experienced a collision without any micro-slot shift in its previous frame. Here, the new frame size as shown in the figure reduces by 21.05% because of slot defragmentation.

Once a node finds a stable micro-slot, it piggybacks control information to all its one-hop neighbors indicating that it is stable. In addition, each node also piggybacks the information indicating the number of micro-slots it has shifted ($\mu\_shift$) to find its stable position. Thus, a node $i$ knows that its one-hop neighbors have found the stable micro-slots and it computes the new frame size ($\widehat{F}$) from the $\mu\_shift$ values from its neighbors $j$ ($\forall j \in$ one-hop neighbors of $i$) using Eqn. (2) as follows.

$$\widehat{F}_i(t) = \widehat{F}_i(t-1) - \max\{\mu\_shift_i(t), \mu\_shift_j(t)\} \quad (2)$$

The pseudo code logic for slot defragmentation (DDSB) executed by each node $i$ is given in Algorithm 1. The overall working model of the entire framework can be understood from Algorithm 2. In short, each node independently learns to find a collision-free transmission schedule using MAB. Post MAB convergence, each node executes the DDSB mechanism periodically to remove any kind of bandwidth usage inefficiency because of time asynchronization or topology change.

```
1: Initialize: μ_shift_i = 0, c_i = 0    // μ_shift_i: Number of micro-
       slot that node i has shifted; c_i: Status of the micro-slot search
       (1, if search is complete, else, 0)
2: If (! Tx in the beginning of frame), do:
3:     Shift to previous micro-slot
4:     μ_shift_i ++
5:     If (Collision ==TRUE):
6:         Check action in the previous frame a(t−1)
7:         If (μ_i(t) > μ_i(t − 1)):
8:             Shift to next two micro-slots
9:             If (Collision ==TRUE):
10:                Shift to previous micro-slot
11:        End If
12:        Else If (μ_i(t) < μ_i(t − 1)):
13:            Shift to next two micro-slots
14:        End If
15:        Set c_i = 1
16:        Piggyback c_i, μ_shift_i
17:        If (c_j == 1 (∀j ∈ one−hop neighbor))
18:            F_shrun(t) = max{μ_shift_i(t), μ_shift_j(t)}
19:            If (F_shrunk(t) == F_shrunk(t − 1)):
20:                Frame Size ← Frame Size − F_shrunk
21:                μ_i(t) = μ_i(t − 1) − F_shrunk
22:                Ignore all collisions
23:            End If
24:        Else:
25:            Set c_i = 1
26:            Piggyback c_i, μ_shift_i
27:            Check the value of c_j, ∀j ∈ one−hop neighbor
28:            If c_j == 1 (∀j ∈ one−hop neighbor)
29:                Find new frame size:
30:                F_shrunk(t) = max{μ_shift_i(t), μ_shift_j(t)}
31:                Frame Size ← Frame Size − F_shrunk
32:                μ_i(t) = μ_i(t − 1) − F_shrunk
33:            End If
34:        End If
35:    End If
36:    If (Collision ==TRUE):
37:        Piggyback adj_i=1      // Control information to neighbors
38:        Shift to previous micro-slot
39:        If (Collision ==FALSE):
40:            Piggyback adj_i = 0
41:        Else:
42:            Shift to next two micro-slots
43:            If (Collision ==FALSE):
44:                Piggyback adj_i = 0
45:            Else:
46:                Shift to previous micro-slot
47:                Piggyback adj_i = 0
48:            End If
49:        End If
50: End If
```

**Algorithm. 1.** DDSB Operation

If a node experiences a collision post MAB convergence, it indicates that there is a new node entering the network and it adjusts its transmission schedule accordingly by increasing the frame size to incorporate the new node. The new node gets to know about the frame size from its neighboring node by piggybacking and it always transmits at the last slot of the frame. This behavior leads to the Decentralized Adaptive TDMA Bandwidth Utilization (DATBU) Protocol executed by each node $i$ independently

1. **Input**: Initial Frame Size ($F_0$)
2. **Initialize**: $t = 0$, $p$   // $p$ is the periodicity for checking exit of a node
3. **While (1)**:
4.     **While (**Convergence==FALSE**)**:
5.         Select transmission slot or mini-slot based on MAB learning policy
6.         Observe reward $R_i$ and update MAB arm value
7.         Check for convergence
8.         **Execute** DDSB Operation (Algorithm 1)
9.         **If** $MOD(t, p) == 0$:
10.            Shift to previous micro-slot
11.            Check Collision
12.            **If** (Collision ! =TRUE):
13.                GO TO Step **9**
14.            **End If**
15.        **End If**
16.        **If** (Collision ==TRUE):
17.            Set $Frame\ Size \leftarrow Frame\ Size + 1$
18.        **End If**
19.        $t++$

**Algorithm. 2.** DATBU Protocol

## VI. EXPERIMENTS AND RESULTS

The experiments with DATBU protocol are performed in a MAC layer simulator with embedded learning components. The simulation kernel performs scheduling in terms of packet generation, transmissions, and receptions.

The ability of the nodes to independently learn using Multi-Armed Bandits, as the first step of the DATBU MAC logic, to find a collision-free slot in a time-synchronized 6-nodes fully connected topology is shown in Fig. 2 (a). The value of the frame-scaling factor $K$ here is set to be 1. It is observed that although there are collisions in the beginning, however, over the time, each node learns to select a transmission slot so that there is no collision post convergence. The ability of the learning framework to find transmission schedule for networks without time synchronization is shown in Fig. 2 (b). The MAB convergence for a 3-nodes fully connected network with constant data rate $\lambda = 1$ packet/frame and $K = \frac{4}{3}$ is shown in the figure. Packet transmission by the nodes with node 1's frame as the frame of reference is plotted. Note that the frames of nodes 2 and 3 lag the frame of node 1 by $0.4\ \tau$ and $0.75\ \tau$ respectively, where $\tau$ represents packet duration. Like the time-synchronized scenario, there are overlapped packet transmissions among the nodes initially. But, over time, all the nodes learn to pick transmission times independently so that no overlapped transmissions take place, and hence collisions are avoided. From the figure, it can be observed that, post convergence, there are unused temporal gaps in the TDMA frame. This is because of the requirement of extra mini-slot for TDMA operation in networks without time synchronization. This bandwidth redundancy can be mitigated using the proposed DDSB operation, as the second step of the DATBU protocol, explained in Section V.

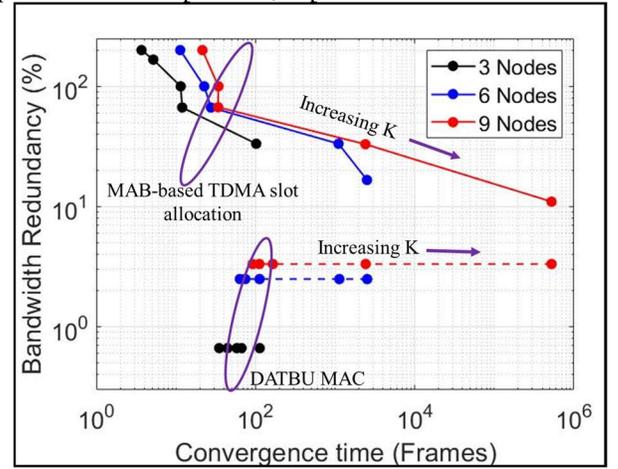

Fig. 3 Convergence speed- Bandwidth redundancy trade-off reduction using DATBU-MAC protocol

The ability of the DDSB operation to reduce bandwidth redundancy in the absence of network time synchronization is tested on a 12-nodes partially connected topology with $K = 2$ (Fig. 2 (c)). It is observed that the excess bandwidth after slot defragmentation reduced from 100% to 15.11% in that topology. The significance of this framework in time-asynchronous fully-connected networks ($K = 2$) can be observed from Bandwidth Usage Efficiency ($BUE = \frac{D}{T} \times 100\%$) in Fig. 2 (d)., where $D$ is the actual duration within a frame of duration $T$

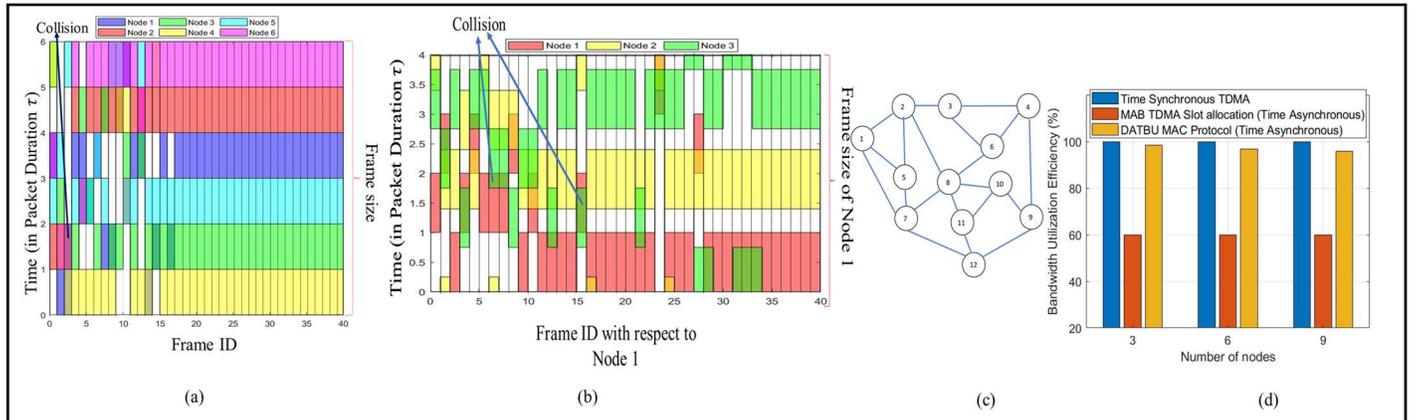

Fig. 2: MAB slot allocation in (a) time-synchronized network (b) time asynchronous network (c) Partially connected topology, (d) Bandwidth utilization efficiency achieved by the proposed framework

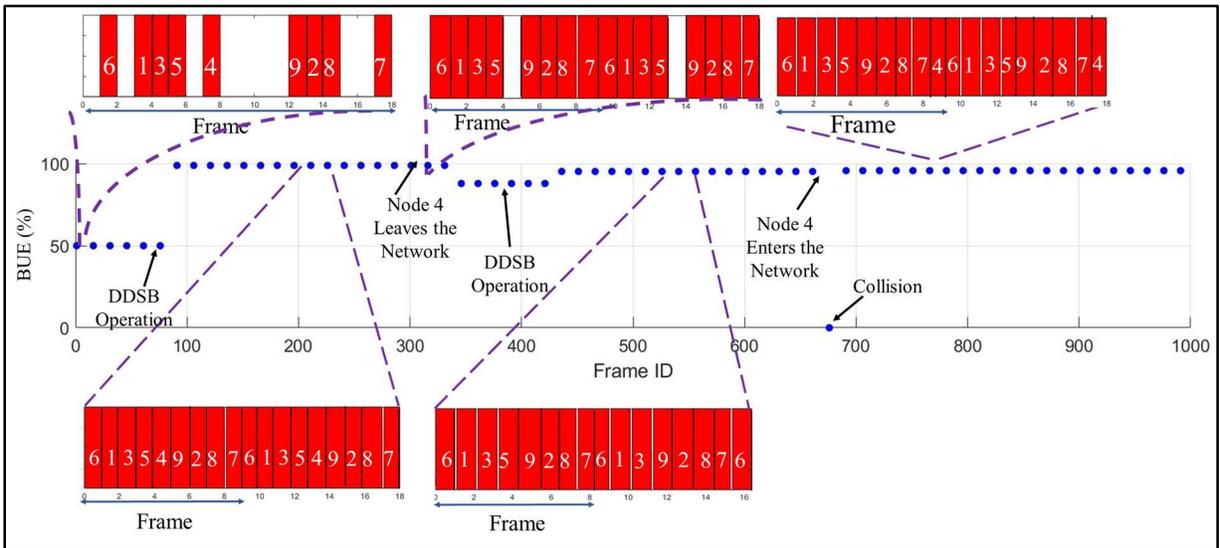
Fig. 4: Adaptability of DATBU protocol to dynamic topologies

that is used for packet transmission. This plot shows three different scenarios: time synchronous TDMA (benchmark), time asynchronous slot allocation by MAB and proposed DATBU protocol without time synchronization. It can be seen that the proposed mechanism achieves $SUE \approx 95\%$ that is close to the time synchronous TDMA. It also shows the significance of the DDSB operation post MAB convergence that improves the bandwidth utilization efficiency $\approx 95\%$.

As mentioned in Section V, there exists a trade-off between convergence time and bandwidth redundancy for MAB-based TDMA (mini)slot allocation. This can be visualized from the solid lines in Fig. 3. This trade-off can be ameliorated using the proposed DDSB mechanism and is demonstrated by the dotted lines in the figure. This is shown to hold for networks of different sizes. To summarize, for a given network and for a given tolerance of bandwidth redundancy, using DDSB helps faster convergence of the learning framework by allowing the network designer to use a larger value of frame scaling factor ($K$). Faster convergence is required, because the nodes cannot afford to lose packets by taking random actions for a long time.

The adaptability of the proposed MAC protocol to topology changes is shown in Fig. 4. The number on each packet in the figure indicates the transmitter of the packet. This figure shows the functioning of the protocol post MAB convergence in a 9 nodes-fully connected topology in a time-synchronized network. It is observed that after DDSB converges, Bandwidth Usage Efficiency ($BUE$) goes to 98%. Moreover, the protocol can adapt to topology changes when node 4 fails to operate. There is drop in BUE when node 4 leaves the network, and BUE is recovered by the adaptive nature of the protocol. Moreover, the protocol can adapt to scenarios of node addition to the network.

## VII. SUMMARY AND CONCLUSIONS

In this paper, a TDMA MAC protocol DATBU is developed for efficient bandwidth utilization in wireless sensor and IoT networks. The proposed protocol has two primary components: an MAB learning-based slot allocation and a slot defragmentation operation called DDSB mechanism for improving bandwidth usage efficiency. The notable feature of this concept is that it is completely decentralized and is suitable for networks without time synchronization capability. Besides managing the trade-off between learning convergence time and bandwidth redundancy, this protocol makes the network adaptive to topological changes. Future direction of this research includes exploring other access performance parameters of the protocol, such as, end-to-end delay and energy efficiency.